\documentclass[reprint,aps,prc,twocolumn,superscriptaddress,floatfix,10pt]{revtex4-2}

\usepackage{bm}
\usepackage{dcolumn}
\usepackage{amsthm}
\usepackage{amssymb}
\usepackage{amsmath}
\usepackage{graphicx}
\usepackage{xcolor}
\usepackage{color}
\usepackage{fix-cm}
\usepackage{mathptmx} 
\usepackage[T1]{fontenc}
\usepackage[colorlinks,allcolors=blue]{hyperref}
\usepackage[utf8]{inputenc}

\usepackage{CJK}

\makeatletter
\def\NAT@def@citea{\def\@citea{\NAT@separator}}
\makeatother

\begin{document}
\begin{CJK*}{UTF8}{}

\title{Microscopic study on fusion reactions $^{40,48}\mathrm{Ca}+{}^{78}\mathrm{Ni}$ and the effect of tensor force}
\author{Xiang-Xiang Sun~\CJKfamily{gbsn}(孙向向)}
\affiliation{School of Nuclear Science and Technology,
 University of Chinese Academy of Sciences,
 Beijing 100049, China}
\author{Lu Guo~\CJKfamily{gbsn}(郭璐)}
\email{luguo@ucas.ac.cn}
\affiliation{School of Nuclear Science and Technology,
University of Chinese Academy of Sciences,
Beijing 100049, China}
\affiliation{CAS Key Laboratory of Theoretical Physics,
              Institute of Theoretical Physics,
              Chinese Academy of Sciences,
              Beijing 100190, China}
\author{A. S. Umar}
\affiliation{Department of Physics and Astronomy,
            Vanderbilt University,
            Nashville, TN 37235, USA}

\begin{abstract}
We provide a microscopic description of the fusion reactions between
$^{40,48}$Ca and $^{78}$Ni.
The internuclear potentials are obtained using the density-constrained (DC)
time-dependent Hartree-Fock (TDHF) approach and fusion
cross sections are calculated via the incoming wave boundary condition method.
By performing DC-TDHF calculations at several selected incident
energies,
the internuclear potentials for both systems are obtained and
the energy-dependence of fusion barrier are revealed.
The influence of tensor force on internuclear potentials of
$^{48}\mathrm{Ca}+{}^{78}\mathrm{Ni}$ is more obvious than those of $^{40}\mathrm{Ca}+{}^{78}\mathrm{Ni}$.
By comparing the calculated fusion cross sections
between $^{40}\mathrm{Ca}+{}^{78}\mathrm{Ni}$ and $^{48}\mathrm{Ca}+{}^{78}\mathrm{Ni}$,
an interesting enhancement of sub-barrier fusion cross
sections for the former system is found,
which can be explained by
the narrow width of internuclear potential for
$^{40}\mathrm{Ca}+{}^{78}\mathrm{Ni}$ while the barrier heights
and positions are very close to each other.
The tensor force suppresses the sub-barrier fusion cross sections of both two systems.
\end{abstract}
\maketitle
\end{CJK*}

\section{Introduction}
Heavy-ion fusion reactions are of particular
importance to extend the nuclear chart
and for the synthesis
of heavy and superheavy elements~\cite{liang2005,adamian2020}.
The development of modern radioactive-ion-beam facilities have greatly broadened
our ability to explore the relation between nuclear structure
and reaction mechanism~\cite{back2014,jiang2021},
especially via fusion reactions of exotic nuclei,
including weakly bound or halo nuclei
and nuclei with a large neutron excess.

The nucleus $^{78}$Ni is found to be doubly magic with a large neutron excess ($N-Z$).
After its discovery~\cite{engelmann1995},
there have been many theoretical and experimental investigations
on its structure properties including shell closures
at $Z=28$ and $N=50$~\cite{hakala2008,sahin2017,olivier2017,welker2017},
half-life~\cite{hosmer2005,xu2014,niu2018},
energy spectra~\cite{mazzocchi2005,hagen2016b,taniuchi2019},
and shape coexistence~\cite{nowacki2016,delafosse2018}.
In a recent investigation on Ni isotopes~\cite{brink2018},
it has been shown that the tensor part of the Skyrme energy density functional (EDF) significantly affects
the spin-orbit splitting of the proton $1f$ orbit, which
may explain the endurance of magicity far from the stability valley.
This in turn further solidifies the durability of the Skyrme EDF in reproducing
the observed shell effects.

But up to now, there has been no theoretical investigation of
fusion reactions with $^{78}$Ni as reactants, particularly also incorporating the tensor force.
To study the influence of neutron excess on fusion,
doubly magic nuclei $^{40,48}$Ca have been widely used as reactants
to study reactions,
such as $^{40,48}\mathrm{Ca}+{}^{90,96}\mathrm{Zr}$~\cite{timmers1997,timmers1998,stefanini2006,stefanini2014},
$^{40,48}\mathrm{Ca}+{}^{124,132}\mathrm{Sn}$~\cite{scarlassara2000,kolata2012},
various Ca$+$Ca systems~\cite{jiang2010,montagnoli2012}, and $^{40}\mathrm{Ca}+{}^{58,64}\mathrm{Ni}$~\cite{bourgin2014}, among others.
It has been shown that the tensor force also affects the heavy-ion
collision process~\cite{fracasso2012,dai2014a,stevenson2016},
the potential barrier, and fusion cross sections~\cite{guo2018,guo2018b,godbey2019c}.
Therefore, it is interesting to study the effects of
tensor force for the reactions $^{40,48}\mathrm{Ca}+{}^{78}\mathrm{Ni}$.
To that end, the purpose of the present investigation is
to predict the fusion cross sections for $^{40,48}\mathrm{Ca}+{}^{78}\mathrm{Ni}$
and study the impact of the tensor force on above and below barrier collisions.
Study of such reactions may become feasible at modern radioactive-ion-beam facilities in the future.

Most of theoretical approaches used to study
the fusion cross sections at
both above and below barrier energies
have similar starting point,
the ion-ion potential.
There are mainly two approaches to determine the ion-ion internuclear potential
to study fusion reactions:
phenomenological ones~\cite{bass1974,randrup1978a,satchler1979,adamian2004,swiatecki2005,feng2006,wang2012,zhu2014b,zagrebaev2015,bao2016,wang2017}
and (semi)microscopic ones~\cite{brueckner1968,diaz-torres2001b,moller2004,guo2004,guo2005,umar2006b,wang2006,misicu2007,diaz-torres2007a,washiyama2008,simenel2013a,wen2013,simenel2017}.
Although phenomenological models have been successfully
applied to study many aspects of reactions data,
their information content is limited due to several
adjustable parameters and the absence of dynamics to formulate potentials, such as the Bass model~\cite{bass1974},
the proximity potential~\cite{seiwert1984,randrup1978a},
the double-folding potential~\cite{satchler1979},
and driven potential from dinuclear system model~\cite{adamian2004}.
Fusion process is particularly complex
and the cross section is influenced by many underlying quantal effects.
For this reason, it is preferable to use a microscopic model that
incorporates the nucleonic
degrees of freedom so that
the nuclear shell structure and the
dynamical effects of the reaction system can be included on the same footing
for a more reliable prediction.

The time-dependent Hartree-Fock (TDHF) approach with
the mean-field approximation,
has been successfully applied to study many aspects of
low-energy heavy-ion collisions
by calculating
the wave functions of nucleons in the three-dimensional grids,
and considering the structure information
of the entrance channel simultaneously
(see Refs.~\cite{simenel2012,nakatsukasa2016,simenel2018,stevenson2019,sekizawa2019} and references therein).
Because the TDHF theory describes the collective motion in a
semiclassical way,
the quantum tunneling of the many-body wave function is not included.
Therefore, the TDHF theory cannot be directly used to describe sub-barrier fusion.
The fusion cross sections at above and below barrier energies
are usually obtained by solving the Schr\"{o}dinger
equation with ion-ion potentials deduced from TDHF calculations.
Several techniques have been developed to obtain
internuclear potential within the framework of TDHF, such as
frozen HF~\cite{guo2012,bourgin2016},
density-constrained (DC) TDHF~\cite{umar2006b},
density-constrained frozen HF~\cite{simenel2017,umar2021},
dissipative-dynamics TDHF~\cite{washiyama2008},
and the Thomas-Fermi approximation~\cite{wang2006}.
Among them, the potentials from the DC-TDHF approach
can naturally incorporate all dynamic effects including nucleon transfer,
neck formation, internal excitations, deformation effects,
and manifestation from the time evolution of the initial configuration.
It has been shown in many studies that the
calculated
fusion cross sections based on the ion-ion potentials with the
DC-TDHF approach are generally in good agreement with measurements~\cite{umar2006b,umar2006d,umar2008b,umar2009b,umar2009a,umar2010c,oberacker2010,keser2012,umar2012a,simenel2013a,simenel2013b,oberacker2013,umar2014a,umar2015c,umar2016,guo2018b,guo2018,godbey2019c,godbey2019b}.

This article is organized as follows.
In Sec.~\ref{Sec2},
we show the main theoretical formulation of the DC-TDHF
approach.
Section~\ref{Sec3} presents the calculational details and the discussion of results.
A summary is provided in Sec.~\ref{Sec4}.

\section{Theoretical framework}
\label{Sec2}
The theoretical framework to calculate the
fusion cross section based on the ion-ion potential
with the DC-TDHF approach has been presented explicitly in
Refs.~\cite{umar2006b,umar2009b,back2014,simenel2018}.
Here we introduce the main formulation of
this approach for the convenience of discussion.
In the TDHF approach, the wave function of the many-body
system is approximated as a single Slater determinant composed
of single-particle states $\phi_i(\bm r)$.
With the mean-field approximation,
the time evolution of the many-body wave function in the three-dimensional space
can be obtained by solving the coupled
nonlinear equations for the evolution of the single-particle states
\begin{equation}
  i\hbar \frac{\partial}{\partial t} \phi_i(\bm r,t)
  = h\phi_i(\bm r,t)\;,\qquad i=1,\cdots, A\;,
\end{equation}
where $h$ is the single-particle Hamiltonian
obtained from the Skyrme effective interaction
including the contributions from time-odd and tensor components.

Since the TDHF approach
does not include the quantum
tunneling of the many-body wave function,
it cannot be
directly applied to study sub-barrier
fusion reactions.
Currently, the fusion reaction is usually treated as
a quantum tunneling through an ion-ion potential
in the center-of-mass frame.
In the DC-TDHF approach,
to extract this internuclear potential
at certain times during the dynamic
evolution,
the instantaneous TDHF density $\rho(\bm r,t)$
is used to get the static HF minimum energy~\cite{umar2006b}.
This many-body state,
$|\Psi_\mathrm{DC}\rangle$, corresponding to static HF
variation process can be obtained by constraining
the total density to be equal to the TDHF instantaneous density,
corresponding to the internuclear separation, $R(t)$
\begin{equation}
  \delta \left\langle \Psi_{\mathrm{DC}}
  \left|
  H-\int d^3 r \lambda(\bm r)\rho(\bm r,t)
  \right|
  \Psi_{\mathrm{DC}}\right\rangle =0\;.
\end{equation}
The ion-ion potential is then obtained by subtracting the binding energies
of projectile and target nuclei, $E_\mathrm{P}$ and $E_\mathrm{T}$, respectively
\begin{equation}
  V(R) = \langle \Psi_\mathrm{DC}|H|\Psi_\mathrm{DC}\rangle
  - E_\mathrm{P} - E_\mathrm{T}\;,
\end{equation}
This approach provides a microscopic
internuclear potential starting from
Skyrme interactions and there is no additional parameters.

Similarly, the coordinate-dependent mass, $M(R)$,
can be obtained using the energy conservation
and affects the energy-dependence of potential~\cite{umar2014a}.
For numerical advantages, we use
the reduced mass $\mu$ and transfer
$V(R)$ into a scaled potential $V(\bar R)$ via
a scale transformation
\begin{equation}
d\bar R =\left(\frac{M(R)}{\mu}\right)^{1/2}dR\;.
\end{equation}

Subsequently, the penetration probabilities $T_L(E_\mathrm{c.m.})$,
corresponding to orbital angular momentum $L$, are obtained by solving the
Schr\"{o}dinger equation
\begin{equation}
\left[\frac{-\hbar^{2}}{2 \mu} \frac{d^{2}}{d \bar R^{2}}
+\frac{L(L+1) \hbar^{2}}{2 \mu \bar R^{2}}+V(\bar R)
-E_{\mathrm{c.m.}}\right] \psi(\bar R)=0\;,
\label{Eq4}
\end{equation}
with the incoming wave boundary condition method~\cite{hagino1999}.
The fusion cross sections at energies below and above
the barrier are then calculated as
\begin{equation}
\sigma_\mathrm{fus}\left(E_{\mathrm{c.m.}}\right)
=\frac{\pi\hbar^2}{2\mu E_\mathrm{c.m.}} \sum_{L=0}^{\infty}
(2 L+1) T_{L}\left(E_{\mathrm{c.m.}}\right)\;.
\end{equation}

\section{Results and discussions}
\label{Sec3}
\begin{figure*}[!htbp]
  \centering
  \includegraphics[width=\textwidth]{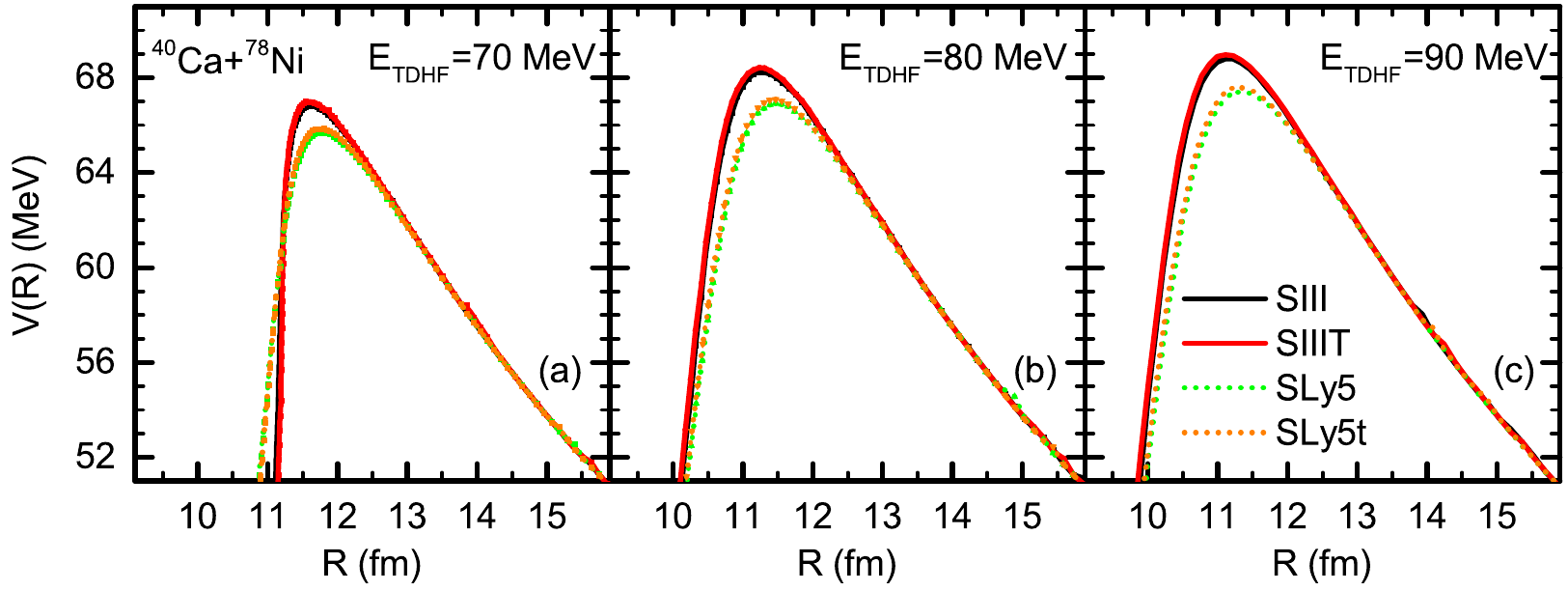}
  \caption{Internuclear potentials obtained from DC-TDHF calculations for $^{40}\mathrm{Ca}+{}^{78}\mathrm{Ni}$
   at $E_\mathrm{TDHF} = 70$ MeV (a), $E_\mathrm{TDHF} = 80$ MeV (b), $E_\mathrm{TDHF} = 90$  MeV (c)
   with density functionals SIII, SIIIT, SLy5, and SLy5t.}
\label{V1}
\end{figure*}

\begin{figure*}[!htbp]
  \centering
  \includegraphics[width=\textwidth]{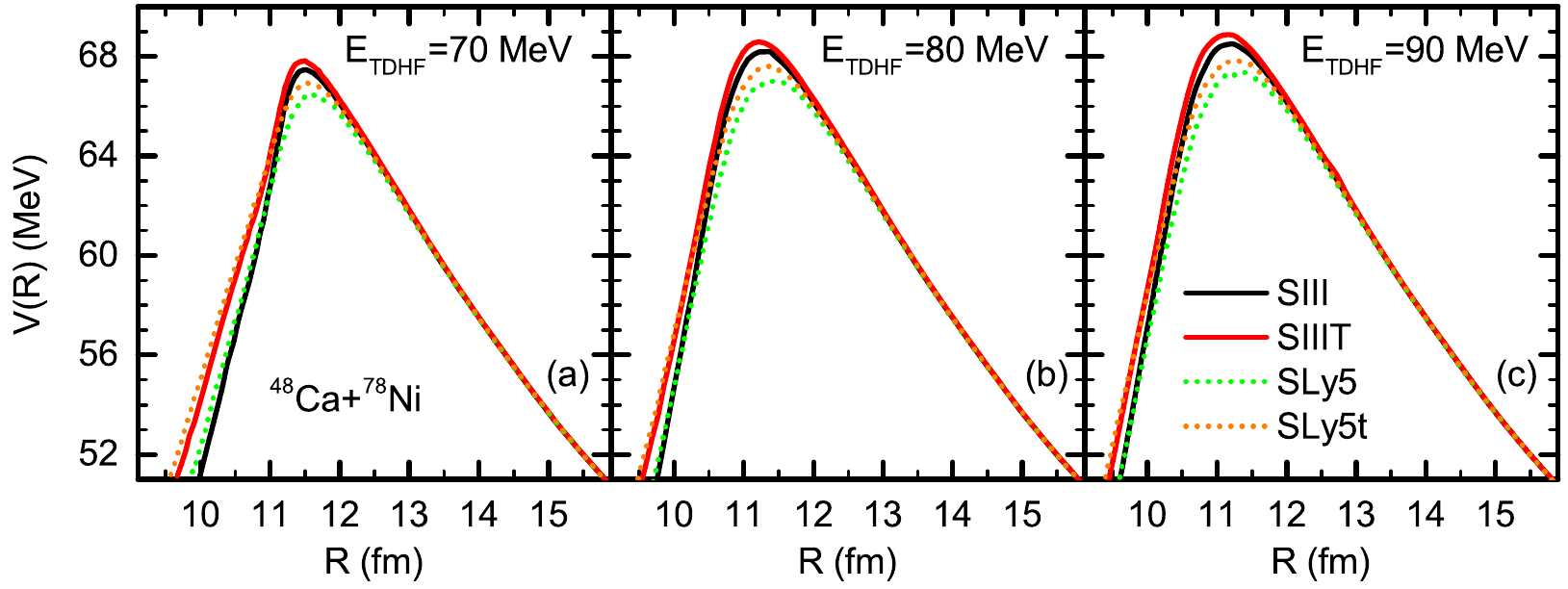}
  \caption{Internuclear potentials obtained from DC-TDHF calculations for $^{48}\mathrm{Ca}+{}^{78}\mathrm{Ni}$
   at $E_\mathrm{TDHF} = 70$ MeV (a), $E_\mathrm{TDHF} = 80$ MeV (b), $E_\mathrm{TDHF} = 90$  MeV (c)
   with density functionals SIII, SIIIT, SLy5, and SLy5t.}
\label{V2}
\end{figure*}

We start our investigations of these reaction systems by performing TDHF
calculations with the modified version of the Sky3D code~\cite{maruhn2014}
that also incorporates the tensor part of the effective interaction. This code
was also used to perform calculations in
Refs.~\cite{dai2014,guo2018,guo2018b,guo2018d,wu2019,wu2020}.
To obtain the ground states of $^{40,48}$Ca and $^{78}$Ni,
the static HF equation are solved on a three-dimensional grid
$28\times 28 \times 28$ fm$^3$.
These three nuclei are all doubly magic and consequently their ground
states are spherical.
Therefore, pairing correlations can be neglected in both static and dynamic calculations.
For the dynamic evolution of central collisions, a three-dimensional
grid with the size of $56 \times 40 \times 40$~fm$^{3}$ is used and the grid spacing in each direction is taken to be 1~fm.
The time step is 0.2~$\mathrm{fm}/c$.
The density constraint calculations
are performed simultaneously
at every 20 time steps (corresponding to 4~fm/$c$ interval).
The initial separation distance of two collision partners is taken to be 20~fm.
The convergence property
in DC-TDHF calculations is as good as that in the
traditional constrained HF with a constraint on a single collective degree of freedom.
All the numerical conditions have been checked for
achieving a good numerical accuracy for all the cases studied here.

In this work, to explore the effects of tensor force on fusion cross section, the Skyrme interactions SLy5~\cite{chabanat1998a} and SLy5t~\cite{colo2007}
(SLy5 plus tensor force) are used.
These two interactions have been used
in many recent investigations with DC-TDHF~\cite{guo2018,guo2018b,godbey2019c,godbey2019b}.
Additionally, in a recent investigation on Ni isotopes~\cite{brink2018},
by comparing the calculations with SIII interaction~\cite{beiner1975} and SIII plus tensor force, denoted by SIIIT,
it has been shown that
the tensor part significantly affects
the spin-orbit splitting of the proton $1f$ orbit
that may explain the survival of magicity far from the stability valley.
Therefore, SIII and SIIIT are also used in the present
study on fusion reactions $^{40,48}\mathrm{Ca}+{}^{78}\mathrm{Ni}$.
The strength of the tensor force is taken to be
the same as that given in
Ref.~\cite{brink2018}.

In Figs.~\ref{V1} and \ref{V2}, we show the ion-ion potentials for $^{40}\mathrm{Ca}+{}^{78}\mathrm{Ni}$
and $^{48}\mathrm{Ca}+{}^{78}\mathrm{Ni}$ from DC-TDHF
calculations at incident energies $E_\mathrm{TDHF}=70$, 80, and
90 MeV with Skyrme interactions
SIII, SIIIT, SLy5, and SLy5t, respectively.
As for $^{40}\mathrm{Ca}+{}^{78}\mathrm{Ni}$, it is found that
the height and position of the fusion barriers calculated by SLy5(t) and SIII(T)
differ by about 1.5~MeV and 0.3~fm,
respectively.
Since the $N(Z)=20$ shell is spin-saturated while the shells with
28 and 50 are not, the inclusion of the tensor force
slightly influences the
height and position of the barriers for $^{40}\mathrm{Ca}+{}^{78}\mathrm{Ni}$.
Although the heights of the barriers increase slowly with
the incident energy in DC-TDHF calculations,
it is not so dramatic when compared with those in heavy or superheavy systems,
such as $^{48}\mathrm{Ca}+{}^{238}\mathrm{U}$~\cite{umar2010a}.
For the $^{48}\mathrm{Ca}+{}^{78}\mathrm{Ni}$ system
the effect of the tensor force on barrier heights are
more pronounced than those for $^{40}\mathrm{Ca}+{}^{78}\mathrm{Ni}$.
The inner part of the potential,
which affects the sub-barrier fusion strongly,
shows a more significant change due to the
tensor force in $^{48}\mathrm{Ca}+{}^{78}\mathrm{Ni}$ than
$^{40}\mathrm{Ca}+{}^{78}\mathrm{Ni}$.
This may be partially due to the fact that $N=28$ shell of $^{48}$Ca is
spin-unsaturated
while the $N(Z)=20$ shell of $^{40}$Ca are spin-saturated.
The effects of tensor force on the internuclear
potentials of $^{40,48}\mathrm{Ca}+{}^{78}\mathrm{Ni}$ is similar with the cases of $^{40,48}\mathrm{Ca}+{}^{48}\mathrm{Ca}$ shown in
Ref.~\cite{guo2018b},
in which the effects of tensor force have been
discussed explicitly and it is shown
that the tensor force influences internuclear potentials
with spin-unsaturated systems rather than spin-saturated ones.
We also note the more pronounced difference in the barrier widths
for the two systems. The effect of this on the fusion cross-sections
will be discussed below.

\begin{figure}[!htbp]
  \centering
  \includegraphics[width=8.6cm]{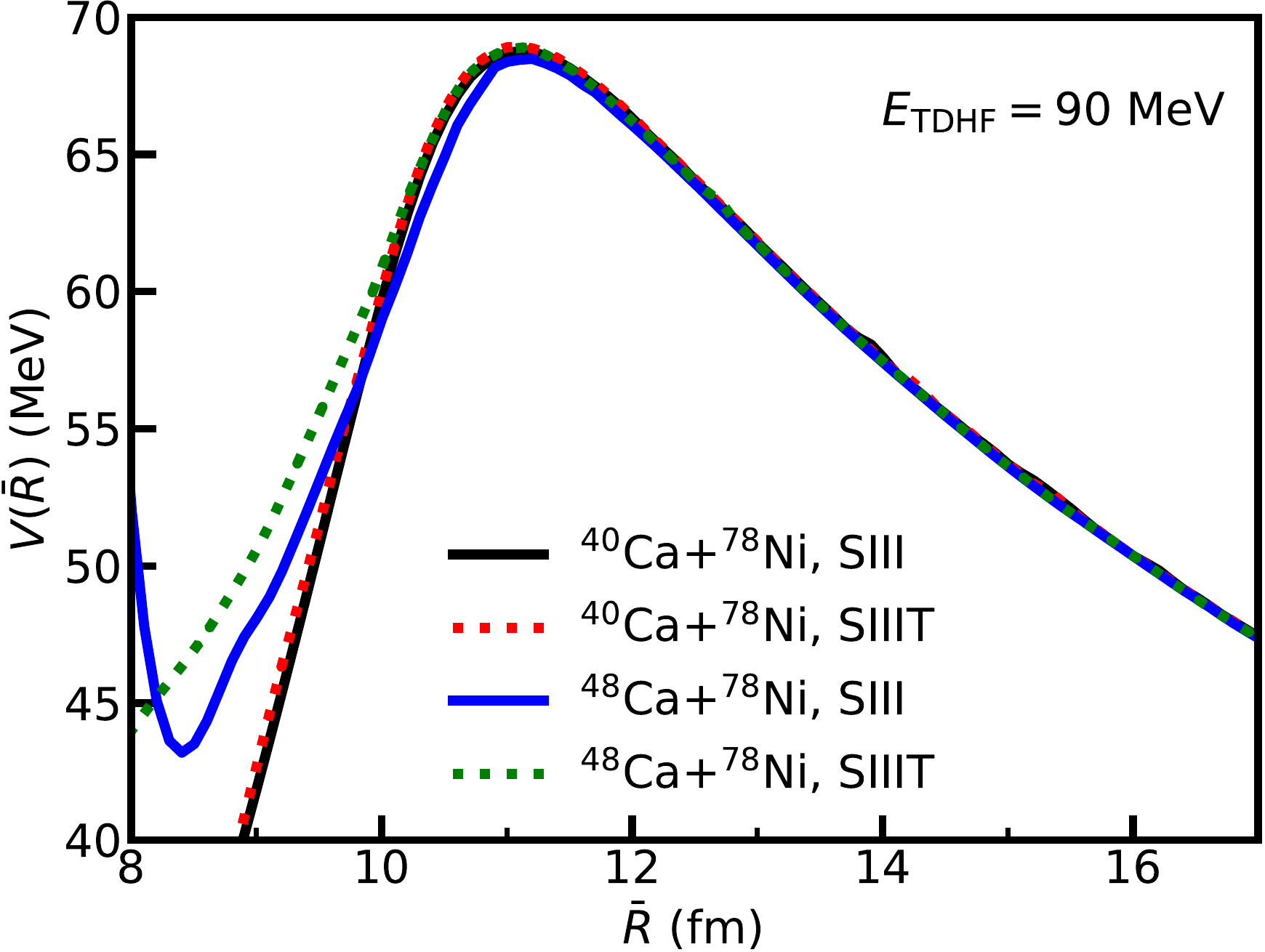}
  \caption{Transformed internuclear potentials for $^{40}\mathrm{Ca}+{}^{78}\mathrm{Ni}$ and $^{48}\mathrm{Ca}+{}^{78}\mathrm{Ni}$
   at $E_\mathrm{TDHF} = 90$  MeV
   with density functionals SIII and SIIIT.}
\label{V3}
\end{figure}

\begin{figure}[!htbp]
  \centering
  \includegraphics[width=8.6cm]{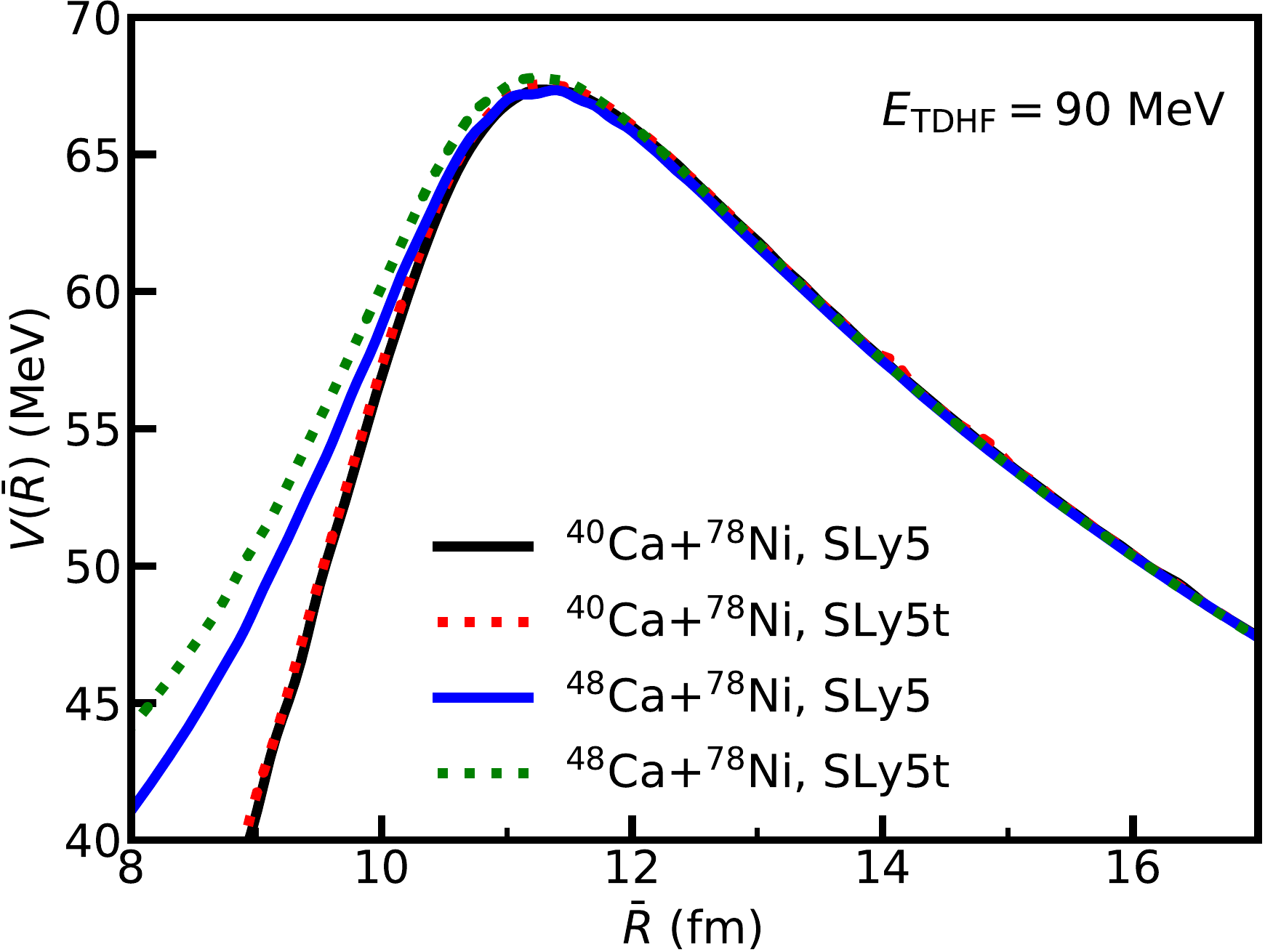}
  \caption{Transformed internuclear potentials for $^{40}\mathrm{Ca}+{}^{78}\mathrm{Ni}$ and $^{48}\mathrm{Ca}+{}^{78}\mathrm{Ni}$
   at $E_\mathrm{TDHF} = 90$  MeV
   with density functionals SLy5 and SLy5t.}
\label{V4}
\end{figure}

The fusion cross sections are calculated by using the
transformed ion-ion potentials with the reduced mass $\mu$.
Before displaying the fusion
cross sections of two reaction systems, we make
a comparison of transformed internuclear potentials $V(\bar R)$
of $^{40}\mathrm{Ca}+{}^{78}\mathrm{Ni}$ with $^{48}\mathrm{Ca}+{}^{78}\mathrm{Ni}$.
Since the energy dependence of potentials from DC-TDHF
calculations for both reactions are relatively weak,
we show
the transformed internuclear
potentials with SIII and SIIIT
effective interactions at the incident c.m. energy $E_\mathrm{TDHF}=90$~MeV in Fig.~\ref{V3}.
As a result of the scale transformation, the inner part
of the potentials are broadened while the outer region of
the potential barrier
are unchanged. This is due to the fact that the
coordinate-dependent mass changes the
interior region of the potential barrier since asymptotically it equals the reduced mass $\mu$
and starts to deviate from this in the interior region.
Similar conclusions can also be obtained
for the potentials calculated with SLy5 and SLy5t, which are shown in Fig.~\ref{V4}.
In addition, the difference caused by the tensor force on the inner region is
enlarged after the scale transformation, especially for the reaction system $^{48}\mathrm{Ca}+{}^{78}\mathrm{Ni}$.
From Figs.~\ref{V3} and \ref{V4}, we see that the position
and height of the barriers for $^{40}\mathrm{Ca}+{}^{78}\mathrm{Ni}$
and $^{48}\mathrm{Ca}+{}^{78}\mathrm{Ni}$ are very close to each other.

After obtaining the potentials, the fusion cross sections can be calculated for all energies
$E_\mathrm{c.m.}$ in the center-of-mass frame.
Usually, if the potentials are strongly dependent on the incident energy $E_\mathrm{TDHF}$
in TDHF simulations, they should
be applied for $E_\mathrm{c.m.}$ intervals close to
a given $E_\mathrm{TDHF}$ and averaged over an energy interval~\cite{umar2014a,oberacker2013}.
For the potential of $^{40}\mathrm{Ca}+{}^{78}\mathrm{Ni}$ with SLy5,
the height and position of the barriers change
from 65.7~MeV and 11.7~fm
at $E_\mathrm{TDHF}=70$~MeV
to 67.4~MeV and 11.3~fm at
$E_\mathrm{TDHF}=90$~MeV.
The potential of $^{48}\mathrm{Ca}+{}^{78}\mathrm{Ni}$ with SLy5
at high energy ($E_\mathrm{TDHF}=90$~MeV)
has a barrier 67.5~MeV located at~11.3~fm,
whereas the potential calculated at low energy
($E_\mathrm{TDHF}=70$~MeV)
has a barrier 66.4~MeV located at 11.6~fm.
For potentials with other effective interactions,
the changes caused by the $E_\mathrm{TDHF}$
are similar to the case of SLy5,
therefore we do not discuss them in detail.
Generally speaking, this energy-dependence of the barrier height
and position for $^{40,48}\mathrm{Ca}+{}^{78}\mathrm{Ni}$ are relatively weak,
which is comparable with that for $^{16}$O+$^{208}$Pb~\cite{umar2009b}.
In present investigation, we calculate the fusion cross section
with the potentials at $E_\mathrm{TDHF}=$ 70, 80, and 90~MeV.

\begin{figure}[!htbp]
  \centering
  \includegraphics[width=8.6cm]{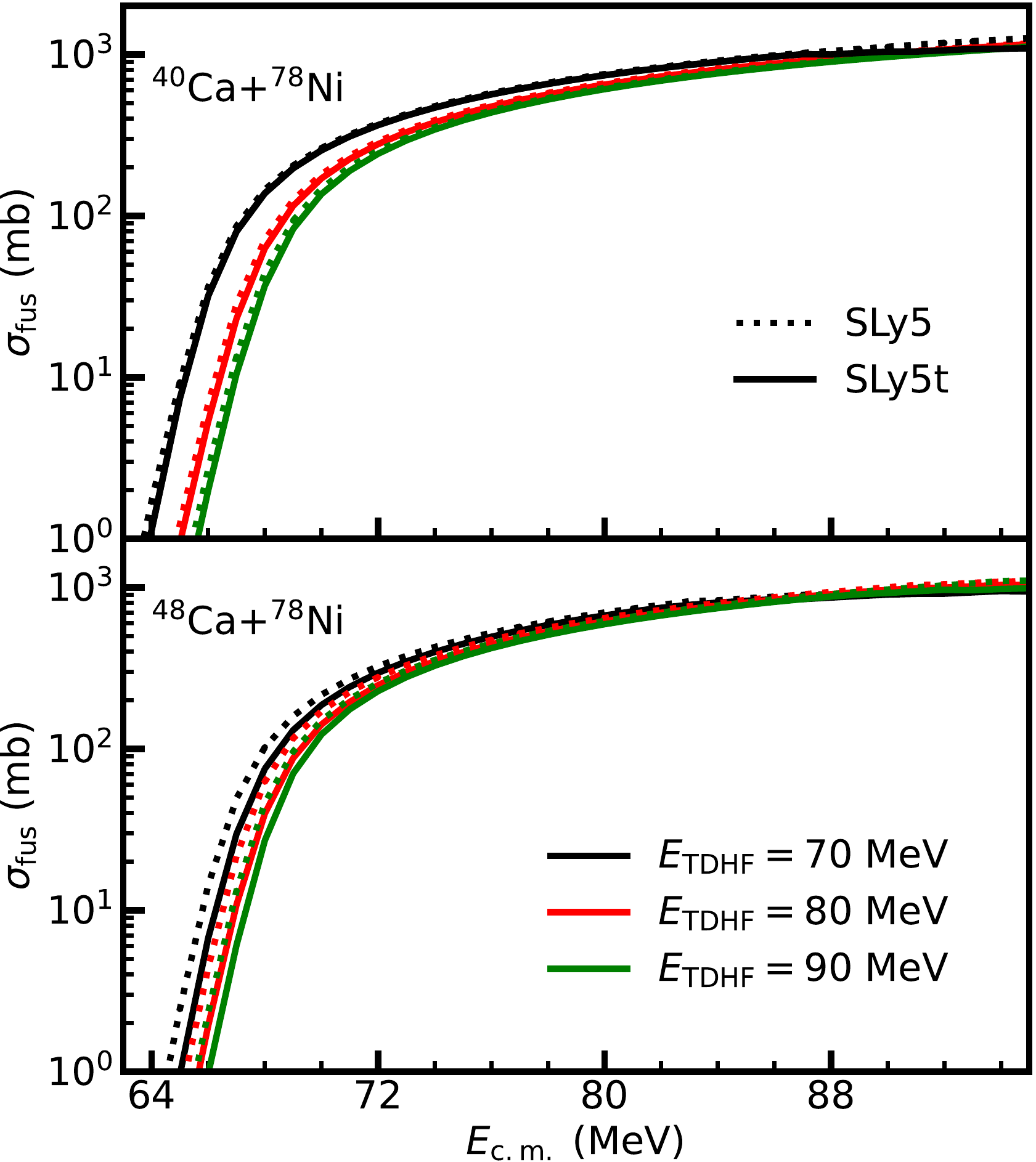}
  \caption{Fusion cross sections for $^{40}\mathrm{Ca}+{}^{78}\mathrm{Ni}$
  (upper panel) and $^{48}\mathrm{Ca}+{}^{78}\mathrm{Ni}$ (bottom panel)
   at $E_\mathrm{TDHF} = 70$~MeV, 80~MeV, and 90~MeV,
   with density functionals SLy5 and SLy5t.}
\label{S1}
\end{figure}

\begin{figure}[!htb]
  \centering
  \includegraphics[width=8.6cm]{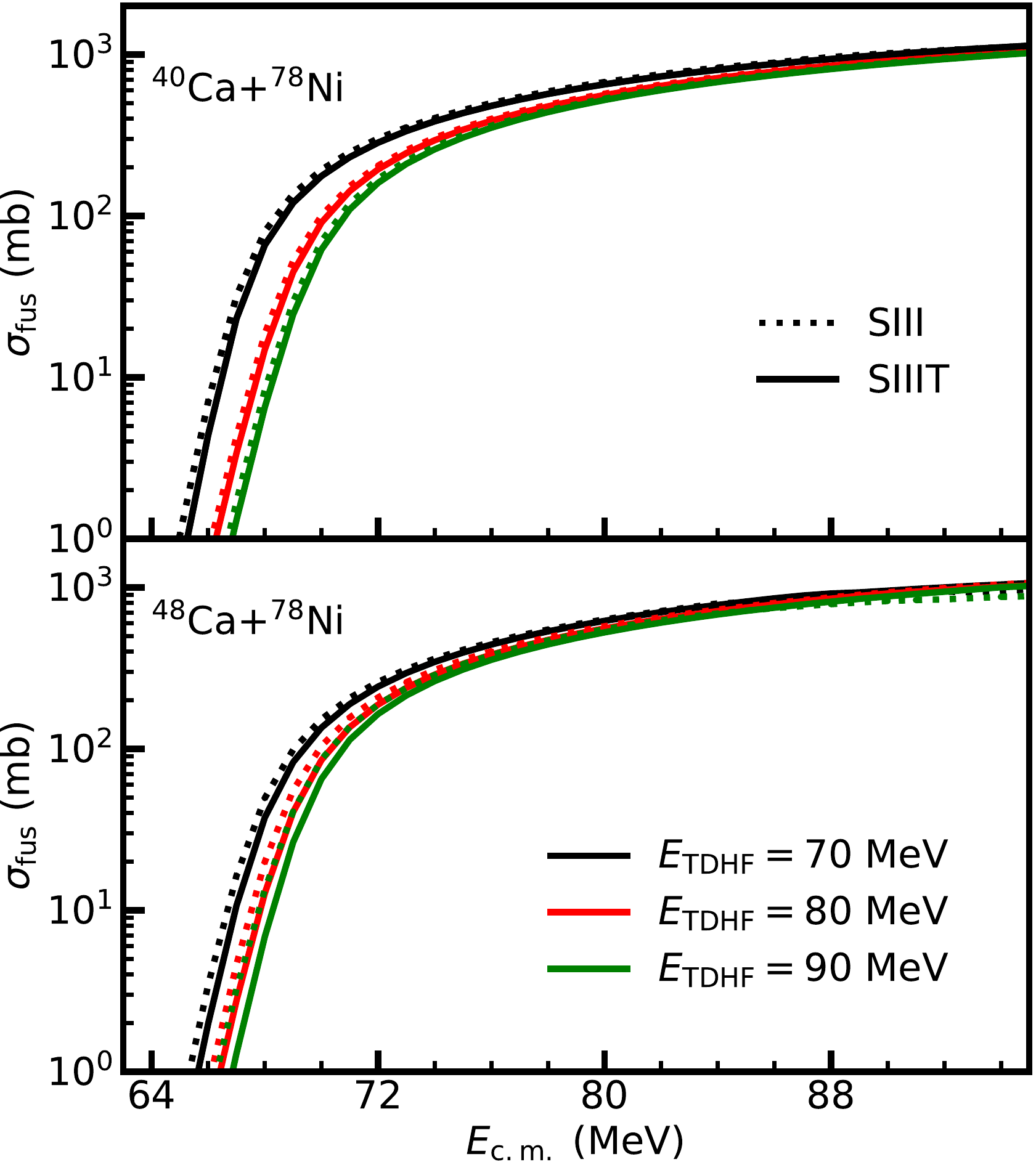}
  \caption{Fusion cross sections for $^{40}\mathrm{Ca}+{}^{78}\mathrm{Ni}$
  (upper panel) and $^{48}\mathrm{Ca}+{}^{78}\mathrm{Ni}$ (bottom panel)
  at $E_\mathrm{TDHF} = 70$~MeV, 80~MeV, and 90~MeV,
  with density functionals SIII and SIIIT.}
\label{S2}
\end{figure}

Figure~\ref{S1} shows the calculated fusion cross sections
for two reaction systems at different $E_\mathrm{TDHF}$ with SLy5 and SLy5t, and those
with SIII and SIIIT are presented in Fig.~\ref{S2}.
First, let us focus on the effects of tensor force.
It is found that the fusion cross sections at
sub-barrier energies are lower after including the
tensor force for both two reaction systems
and the magnitude of
this hindrance for $^{48}\mathrm{Ca}+{}^{78}\mathrm{Ni}$ is slightly
larger than that for $^{40}\mathrm{Ca}+{}^{78}\mathrm{Ni}$.
This is due to the tensor force increasing the potential barrier
and broadening the inner part the potential.
The influence of the tensor force on the fusion cross section
and the conclusions obtained
in the present case are in line with the conclusions of Ref.~\cite{godbey2019c}.

By comparing the fusion cross sections
of two reaction systems,
it is found that at sub-barrier energies,
the fusion cross sections of ${}^{40}\mathrm{Ca}+{}^{78}\mathrm{Ni}$
are larger than those of $^{48}\mathrm{Ca}+{}^{78}\mathrm{Ni}$
by several times, meaning an enhancement of fusion
cross sections at sub-barrier energies for $^{40}\mathrm{Ca}+{}^{78}\mathrm{Ni}$
as compared to the more neutron-rich system $^{48}\mathrm{Ca}+{}^{78}\mathrm{Ni}$.
In our cases, by comparing the internuclear
potentials, one observes that the height and position of the
barriers for both two systems are almost the same,
but the width of $^{40}\mathrm{Ca}+{}^{78}\mathrm{Ni}$ is narrower than that of $^{48}\mathrm{Ca}+{}^{78}\mathrm{Ni}$,
resulting in the enhancement of sub-barrier fusion cross sections.
This situation is similar with the measurements
of $^{132}\mathrm{Sn}+{}^{40}\mathrm{Ca}$ and $^{132}\mathrm{Sn}+{}^{48}\mathrm{Ca}$~\cite{kolata2012}.
The enhancement of sub-barrier fusion cross sections for
$^{132}\mathrm{Sn}+{}^{40}\mathrm{Ca}$ has also been successfully explained by the DC-TDHF calculations~\cite{oberacker2013} and is due to its narrower width of the ion-ion potential.
Additionally, it is should be mentioned that
the transfer channels also play an important role for the sub-barrier fusion
and one can use the particle number projection methodr~\cite{simenel2010} to
estimate the neutron transfer probabilities. But this is beyond the scope of
the present study.

\section{Summary}
\label{Sec4}
Doubly magic nucleus $^{78}$Ni has a very large neutron excess and
its properties are connected with many essential ingredients
of nuclear-structure studies, thus drawing many theoretical and experimental interests.
In this work, we present the first microscopic study of the fusion reactions involving this nucleus.
By using the DC-TDHF approach,
the ion-ion potentials of
$^{40}\mathrm{Ca}+{}^{78}\mathrm{Ni}$ and $^{48}\mathrm{Ca}+{}^{78}\mathrm{Ni}$ are obtained with the only input being the Skyrme effective interactions.
By comparing the internuclear potential calculated by SIII (SLy5)
with SIIIT (SLy5t), we find that the tensor force
increases slightly the potential barrier and broadens the inner part the potential,
though the magnitude differs by the system.
In addition, the inclusion of the tensor force
suppresses the fusion cross sections at sub-barrier region.
More interesting,
it is found that the height and position of the potential
barriers for $^{40}\mathrm{Ca}+{}^{78}\mathrm{Ni}$ and $^{48}\mathrm{Ca}+{}^{78}\mathrm{Ni}$
are very close to each other while the barrier width of
$^{40}\mathrm{Ca}+{}^{78}\mathrm{Ni}$ is more narrower than that of
$^{48}\mathrm{Ca}+{}^{78}\mathrm{Ni}$.
This results in an enhancement of fusion cross sections of $^{40}\mathrm{Ca}+{}^{78}\mathrm{Ni}$
at sub-barrier energies.
The reactions $^{40,48}\mathrm{Ca}+{}^{78}\mathrm{Ni}$ are expected
to be performed at modern radioactive-ion-beam facilities
and the predictions presented in this work provide
a prior theoretical support.

\acknowledgments
We thank Shan-Gui Zhou for helpful discussions.
This work has been supported by
the National Natural Science Foundation of China (Grants No. 11975237,
No. 11575189, and No. 11790325) and
the Strategic Priority Research Program of Chinese Academy of Sciences
(Grant No. XDB34010000 and No. XDPB15) and by the U.S. Department of Energy under grant No.
DE-SC0013847 with Vanderbilt University.
The results described in this paper are obtained on
the High-performance Computing Cluster of ITP-CAS and
the ScGrid of the Supercomputing Center,
Computer Network Information Center of Chinese Academy of Sciences.


%

\end{document}